\documentclass[showkeys,nofootinbib,prd]{revtex4}

\usepackage{amsmath}
\usepackage{amsfonts}
\usepackage{amssymb}
\usepackage{amsthm}
\usepackage{mathtools}
\usepackage{subfigure}
\DeclareFontFamily{U}{mathb}{\hyphenchar\font45}
\DeclareFontShape{U}{mathb}{m}{n}{
      <5> <6> <7> <8> <9> <10> gen * mathb
      <10.95> mathb10 <12> <14.4> <17.28> <20.74> <24.88> mathb12
      }{}
\DeclareSymbolFont{mathb}{U}{mathb}{m}{n}

\DeclareMathSymbol{\Sun}{3}{mathb}{"40}

\allowdisplaybreaks

\begin{document}
\title{Gravitational lensing in a spacetime with extra dimensions}
\author{Mattia Villani}
\affiliation{University of Urbino Carlo Bo, Department of Pure and Applied Sciences (DiSPeA), Via Santa Chiara, 27, Urbino (PU), 61029, Italy}
\email{mattia.villani@uniurb.it}

\begin{abstract}
This is the fourth paper of a series in which we consider the possibility to use cosmological extra dimensions to explain the accelerated expansion of the Universe and the rotation curve of spiral galaxies without introducing dark matter and dark energy. Here we study gravitational lensing in this setting; we derive an expression for time delay and present a modified likelihood for the mass reconstruction procedure which takes into account the effect of the expansion or contraction of the extra dimensions.
\end{abstract}
\maketitle

\section{Introduction}

Since the first years of General Relativity, it is known that a gravitational field can bend light. Over time, gravitational lensing has become an important tool in the study of cosmology because it can map the profile of the mass of the lens, allowing to study the distribution of dark matter in galaxy clusters. There are three types of gravitational lensing: strong, weak, and microlensing. Strong lensing produces multiple distorted images of the same objects (for example, Einstein crosses, see \cite{cross}) or strong deformation of the background object (such as Einstein rings, see, for example, \cite{ring}). Multiple images follow different paths through the Universe and arrive at the observer at different times; this time delay gives the possibility to study the mass distribution of the lens \cite{td2} and cosmological models \cite{td}. Weak lensing produces small modifications in the shape of background objects. The statistical study of these distortions allows the mapping of the mass in the lens \cite{weak}. Microlensing causes an increase in the luminosity of a distant object; it has several applications, for example, the study of MACHOs \cite{macho,macho2} or esoplanets \cite{esoplanet}. The various types of lensing can also be combined; for example, a galaxy can work as a lens producing distortions of another more distant galaxy, and a single star of the former can, for a short time, increase the luminosity of a single star in the latter \cite{combined1,combined2}.

In the recent paper \cite{mio}, we have considered a cosmological model with extra dimensions where the 4d submanifold is described by a FLRW metric with scale factor $a$, while the $n$ extra dimensions are compactified to a hypersphere with time dependent radius $R$ in principle different from the scale factor $a$; the metric of this spacetime is given by
\begin{equation}\label{eq:met}
    ds^2=dt^2-\dfrac{a(t)^2}{1-r^2\,\kappa}dr^2-a(t)^2\,r^2\,d\Omega_2-R(t)^2\,d\Omega_n
\end{equation}
where $d\Omega_i$ is the usual metric of a $i-$dimensional sphere and where $\kappa=\{-1,0,1\}$ is the curvature parameter of the FLRW metric. In that paper, we have shown that the dark energy density is given by
\begin{equation}
    \Omega_{\Lambda}(z)=\dfrac{5}{3}\dfrac{\mathcal{H}^2(z)}{H_0^2}-\dfrac{2\,\mathcal{H}(z)}{H_0}\,\sqrt{-\dfrac{G(z)}{3H_0^2}+\Omega^{Real}_m\,(1+z)^3-\Omega_k\,(1+z)^2+\dfrac{2}{3}\,\dfrac{\mathcal{H}^2(z)}{H_0^2}},
\end{equation}
where $G(z)$ is the Gaussian curvature of the extra dimensions and $\mathcal{H}=\dot{R}/R$ is a Hubble-like parameter that describes the expansion rate of the extra dimensions. At the same time, we have found that we can identify
\begin{equation}
    \Omega^{Meas}_m=\Omega_m^{Real}-\dfrac{G}{3H_0^2}
\end{equation}
as the measured density parameter, while $\Omega_m^{Real}$ is the real value of the mass density, i.e., that the Gaussian curvature reduces the measured (dark and visible) matter density of the Universe. Finally, we have calculated the luminosity distance up to $O(z^2)$ and the acceleration parameter, finding that it is also \emph{dressed} by terms coming from the extra dimensions. 

In paper \cite{mio2}, we have made the hypothesis that locally $R=R(r)$, i.e., that locally the radius is independent from time, but depends on the radial coordinate. We have found that with this modification we could describe the rotation curve of spiral galaxies and obtain the growth of primordial overdensities even without introducing dark matter. 

Here we continue on the study of this spacetime; in particular, we focus on gravitational lensing. In Section \ref{sec:lens}, we calculate the lens equation, lens potential and convergence for thin lenses finding that they are all modified by the presence of extra dimensions. We calculate the time delay in Section \ref{sec:td} and in Section \ref{sec:mrec} we present a mass reconstruction method that take into account the extra dimensions and provide a way to estimate the parameter $\mathcal{H}$, the rate of expansion of the extra dimensions. Finally, in Section \ref{sec:weak}, we provide a framework for the weak lensing and its effect on the CMB.

\section{The lens equation}
\label{sec:lens}

We follow closely \cite{bart}. In our spacetime \eqref{eq:met}, we define a null geodesic with tangent vector $v^i$. The geodesic deviation equation is given by
\begin{equation}
    \nabla^2_kv^i=R(k,v^i)\,k
\end{equation}
where $k$ is the wave vector of the null geodesic and $R$ is the Riemann tensor. Following the discussion in \cite{bart} we can arrive up to his equation (18)
\begin{equation}
    \nabla^2_kv^i=\dfrac{d^2v^i}{d\lambda^2}=\mathcal{R}\,v^i, \qquad \mathcal{R}=-\dfrac{2\pi G}{c^2}\,\rho\,(1+z)^5.
\end{equation}
Then, considering $x^i=v^i/a$, the comoving bundle dimension, and using the comoving radial distance $r$ in place of $\lambda$, we find
\begin{equation}\label{eq:uno}
    \dfrac{d^2\,x^i}{dr^2}=a^2\,\dfrac{d}{d\lambda}\left( (v^i)^\prime\,a-v^i\,a^\prime \right)=a^2\,\left( \dfrac{d^2v^i}{d\lambda^2}a-v^i\,a^{\prime\prime} \right).
\end{equation}
We have
\begin{equation}
    a^\prime=\dfrac{\dot{a}}{ca}, \qquad a^{\prime\prime}=\dfrac{1}{2c^2}\,\dfrac{d}{da}\left( \dfrac{\dot{a}}{a} \right)^2.
\end{equation}
substituting the above expressions in \eqref{eq:uno} and using the definition of the Hubble parameter \cite{mio}
\begin{equation}\label{eq:hubble1}
    \left( \dfrac{\dot{a}}{a} \right)^2\,\dfrac{1}{H_0^2}=\dfrac{\Omega^{Real}_m}{a^3}-\dfrac{\Omega_k}{a^2}-\dfrac{n\,G}{6H_0^2}+\dfrac{n(1+2n)}{6}\dfrac{\mathcal{H}^2}{H_0^2}-\dfrac{n\,\mathcal{H}}{3H_0}\,\sqrt{-\dfrac{n\,G}{3H_0^2}+\dfrac{\Omega^{Real}_m}{a^3}-\dfrac{\Omega_k}{a^2}+\dfrac{n(2+n)}{3}\,\dfrac{\mathcal{H}^2}{H_0^2}},
\end{equation}
we arrive at the equation
\begin{equation}\label{eq:prima}
    \dfrac{d^2\,x^i}{dr^2}=-\kappa\,\left( 1-n\,\dfrac{\mathcal{H}(z)}{H_0\,S(z)} \right)\,x^i-n\,\dfrac{\Omega_m^{Real}\mathcal{H}(z)}{H_0\,S(z)} \,x^i,\qquad S(z)=\sqrt{-\dfrac{n\,G}{3H_0^2}+\dfrac{\Omega^{Real}_m}{a^3}-\dfrac{\Omega_k}{a^2}+\dfrac{n(2+n)}{3}\,\dfrac{\mathcal{H}^2}{H_0^2}}\, \quad a=(1+z)^{-1}
\end{equation}
which reduces to the usual one when $n\rightarrow0$. In the case where there are local perturbations in the mass field, the metric \eqref{eq:met} is modified as follows
\begin{equation}\label{eq:met2}
    ds^2=(1+2\phi)dt^2-(1-2\phi)\left[\dfrac{a(t)^2}{1-r^2\,\kappa}dr^2-a(t)^2d\Omega_2\right]-R(t)^2\,d\Omega_n
\end{equation}
where $\phi$ is the potential of the perturbation and equation \eqref{eq:prima} is modified as follows \cite{bart}
\begin{equation}\label{eq:seconda}
    \left[ \dfrac{d^2}{dr^2} +\kappa\,\left( 1-n\,\dfrac{\mathcal{H}(z)}{H_0\,S(z)} \right)+n\,\dfrac{\Omega_m^{Real}\mathcal{H}(z)}{H_0\,S(z)} \right] \, x^i=-2\,\partial^i\phi.
\end{equation}

In order to solve this equation, one needs the Green's function of the operator on the LHS. It is a complicated matter to solve for this Green's function for general $\mathcal{H}(z)$ and $S(z)$. The redshift can be rewritten in terms of the coordinate $r$. We assume that the leading order in this expansion is sufficient, so we obtain
\begin{equation}
    \dfrac{\mathcal{H}(z)}{S(z)}\approx\dfrac{\mathcal{H}_d}{S_d}
\end{equation}
where $\mathcal{H}_d$ and $S_d$ are the values of the expansion and of the $S$ parameter at the position of the lens. Then, we have
\begin{equation}\label{eq:terza}
    \left[ \dfrac{d^2}{dr^2} + \kappa\,\left( 1-n\,\dfrac{\mathcal{H}_d}{S_d} \right) + \dfrac{3n\,\Omega_m^{Real}}{2}\,\dfrac{\mathcal{H}_d}{S_d} \right] \, x^i=-2\,\partial^i\phi.
\end{equation}
A Green's function for the LHS operator is obtained with the boundary conditions \cite{bart}
\begin{equation}
    \left.x^i\right|_0=0, \qquad \left.\dfrac{dx^i}{dr}\right|_0=\theta^i.
\end{equation}
We find
\begin{equation}
    G(r,r^\prime)=\dfrac{\sin(b\,(r-r^\prime))}{b}\,\Theta(r-r^\prime), \qquad b=\sqrt{\kappa\,\left( 1-\dfrac{\mathcal{H}_d}{S_d} \right) + \Omega_m^{Real}\,\dfrac{\mathcal{H}_d}{s_d}},
\end{equation}
where $\Theta(x)$ is the Heavyside theta function. For a flat spacetime ($\kappa=0$), we have
\begin{equation}\label{eq:green}
    G(r,r^\prime)=\dfrac{\sin\left(\sqrt{n\,\Omega_m^{Real}\,\dfrac{\mathcal{H}_d}{S_d}}\,(r-r^\prime)\right)}{\sqrt{n\,\Omega_m^{Real}\,\dfrac{\mathcal{H}_d}{S_d}}}\,\Theta(r-r^\prime)=f(r,r^\prime)\,\Theta(r-r^\prime),
\end{equation}
which reduces to the usual one for $n=0.$

A solution to equation \eqref{eq:terza} is
\begin{equation}
    \beta^i=\dfrac{x^i}{f(r_s)}=\theta^i-2\int_0^{r_s} dr^\prime \dfrac{f(r_s-r^\prime)}{f(r_s)}\,\partial\phi\left( f(r^\prime)\theta^i,r^\prime \right)
\end{equation}
which is called the lens equation (a WKB approximation has been performed to obtain the above equation, see \cite{bart}). The second term
\begin{equation}\label{eq:defl}
    \alpha^i=2\int_0^{r_s} dr^\prime \dfrac{f(r_s-r^\prime)}{f(r_s)}\,\partial\phi\left( f(r^\prime)\theta^i,r^\prime \right)
\end{equation}
is the deflection angle. The lens potential is given by
\begin{equation}\label{eq:len_pot}
    \psi(\theta^i)=2\int_0^{r_s} dr^\prime \dfrac{f(r_s-r^\prime)}{f(r_s)\,f(r^\prime)}\,\partial\phi\left( f(r^\prime)\theta^i,r^\prime \right).
\end{equation}

From the lens potential, we can define the convergence $K$ and the shear $\gamma_i$:
\begin{align}\label{eq:k}
    K&=\dfrac{1}{2}\, \partial^i\partial_i\,\psi=\dfrac{1}{2}\,\psi^i_i,\\
    \gamma_1&=\dfrac{1}{2}\,(\psi^1_1-\psi^2_2), \quad \gamma_2=\psi_1^2=\psi_2^1.
\end{align}
In general a circular source of unit radius is mapped to an ellipse with semi-major and semi-minor axes given by
\begin{equation}
    A=(1-K-|\gamma|)^{-1}, \qquad B=(1-K+|\gamma|)^{-1}
\end{equation}
which together define the measured ellipticity:
\begin{equation}
    \epsilon=\dfrac{A-B}{A+B}=\dfrac{|\gamma|}{1-K}.
\end{equation}

Given the definitions of the lens potential \eqref{eq:len_pot} and of the convergence \eqref{eq:k}, we find \cite{bart}
\begin{equation}
    K=\dfrac{4\pi G}{c^2}\,\int_0^{r_s}dr^\prime \dfrac{f(r^\prime)f(r_s-r^\prime)}{f(r_s)}\,\rho(f(r^\prime)\theta^i,w^\prime),
\end{equation}
where $\rho$ is the density profile of the lens.

Since lenses are typically much smaller than the distances involved, a thin lens approximation is usually a good one. This is performed by imposing \cite{bart}
\begin{equation}
    \phi=\delta(r-r_d)\,\int dr^\prime \phi=\delta(r-r_d)\,\phi^{(2)},
\end{equation}
where $r_d$ is the distance of the lens from the observer. Then, the lens equation \eqref{eq:defl} becomes
\begin{equation}
    \beta^i=\theta^i-2\,\dfrac{f(r_{ds})f(r_d)}{f(r_s)}\,\partial^i\,\phi^{(2)}, \qquad r_{ds}=r_s-r_d
\end{equation}
while the lens potential and the convergence become, respectively
\begin{align}
    \psi&=2\,\dfrac{f(r_{ds})}{f(r_d)f(r_s)}\,\partial^i\,\phi^{(2)},\\
    K&= \dfrac{4\pi G}{c^2}\,\dfrac{f(r_{ds})f(r_d)}{f(r_s)}\,\Sigma(\theta^i)=\dfrac{\Sigma(\theta^i)}{\Sigma_{cr}},
\end{align}
where $\Sigma(\theta^i)$ is the surface density of the lens and $\Sigma_{cr}$ is called the critical surface mass density.

We now specialize our equations to simple lens profiles.

\subsection{Point mass}
A point mass is described by a density of the form $\rho(\vec{x})=M\,\delta(\vec{x})$. In this case, the convergence and the lens potential are given by
\begin{align}
    K&=\dfrac{4\pi G\,M}{c^2}\,\dfrac{f(r_s-r_d)\,f(r_d)}{f(r_s)}\,\delta(r_d\,\vec{\theta}),\\
    \psi&=\dfrac{4\pi G M}{c^2}\,\dfrac{f(r_s-r_d)}{f(r_s)\,f(r_d)}\,\ln(|\theta|).
\end{align}
From the definition of $f(r)$, equation \eqref{eq:green}, we can assume that $\mathcal{H}_d/H_0\ll1$, which is justified by what we found in \cite{mio}, thus expanding the above equations up the first order in $\mathcal{H}_d$, we find
\begin{align}
    K&=\dfrac{4\pi G\,M}{c^2}\,\dfrac{r_{ds}\,r_d}{r_s}\,\left( 1-\dfrac{n\,\Omega_m^{Real}}{3}\,\dfrac{\mathcal{H}_d}{H_0S_d}\, r_{ds}\,r_d\right)\,\delta(r_d\,\vec{\theta}),\\
    \psi&=\dfrac{4\pi G M}{c^2}\,\dfrac{r_{ds}}{r_s\,r_d}\,\left( 1-\dfrac{n\,\Omega_m^{Real}}{3}\,\dfrac{\mathcal{H}_d}{H_0S_d}\, r_{s}\,r_d\right)\,\ln(|\theta|).
\end{align}
In this approximation, the deflection angle is
\begin{equation}
    \alpha^i=\dfrac{4\pi G M}{c^2}\,\dfrac{r_{ds}}{r_s\,r_d}\,\left( 1-\dfrac{n\,\Omega_m^{Real}}{3}\,\dfrac{\mathcal{H}_d}{H_0S_d}\, r_{s}\,r_d\right)\,\dfrac{\theta^i}{\theta^2}.
\end{equation}

We notice that all the above equations have a correction due to the presence of the extra dimensions. This correction is different for the convergence and for the deflection angle and depends on the distances of the source and of the lens.

\subsection{Galaxy profile}

The density profile of a galaxy is given by the non-singular isothermal sphere \cite{bart}
\begin{equation}
    \rho=\dfrac{\sigma^2}{2\pi G}\,\dfrac{1}{r^2+r_c^2}
\end{equation}
where $\sigma$ is the dispersion and $r_c$ the core radius. With this profile, we obtain the convergence
\begin{equation}
    K=\dfrac{2\pi\sigma}{c^2}\,\dfrac{r_{ds}}{r_s}\,\left( 1- \dfrac{\Omega_m^{Real}}{3}\,\dfrac{\mathcal{H}_d}{H_0S_d}\,r_{ds}\,r_d \right)\,\dfrac{1}{\sqrt{\theta^2+\theta^2_c}},
\end{equation}
while the lens potential is
\begin{equation}
    \psi=\dfrac{4\pi\sigma}{c^2}\dfrac{r_{ds}}{r_s}\,\left( 1-\dfrac{n\,\Omega_m^{Real}}{3}\,\dfrac{\mathcal{H}_d}{H_0S_d}\,r_dr_s \right)\,\left[ \sqrt{\theta^2+\theta^2_c}-\theta_c\,\ln\left( \dfrac{\theta_c+\sqrt{\theta^2+\theta^2_c}}{\theta} \right) \right].
\end{equation}
We see in the above expressions the same corrections we have found in the case of the point mass.

\section{Time delay}
\label{sec:td}
In this Section, we calculate the time delay due the gravitational lensing. It is known that there are two different delays: one due to the different path of light, and one due to the gravitational redshift.

The path delay can be calculated as in \cite{bart} using the cosine theorem, we find
\begin{equation}
    \Delta t_{path}=\dfrac{r_sr_d}{r_{ds}}\,\dfrac{(\theta-\beta)^2}{2c}
\end{equation}
where $\beta$ is given by \eqref{eq:terza}. The gravitational delay is given by the lens potential $\psi$ as follows
\begin{equation}
    \Delta t_{gr}=-\dfrac{1}{c}\,\dfrac{r_sr_d}{r_{ds}}\,\left( 1-\dfrac{n\,\Omega_m^{Real}}{3}\,\dfrac{\mathcal{H}_d}{H_0S_d}\,r_sr_d \right)\,\psi
\end{equation}

Putting them together, we find
\begin{subequations}
\begin{equation}
    \Delta t=\Delta t_{path}+\Delta t_{gr}=\dfrac{1}{c}\,\dfrac{r_sr_d}{r_{ds}}\,\left[ \dfrac{(\theta-\beta)^2}{2}- \psi\,\left( 1-\dfrac{n\,\Omega_m^{Real}}{3}\,\dfrac{\mathcal{H}_d}{H_0S_d}\,r_sr_d \right) \right]=\Delta t_0+\delta_t.
\end{equation}
where
\begin{equation}\label{eq:t0}
    \Delta t_0=\dfrac{1}{c}\,\dfrac{r_sr_d}{r_{ds}}\,\left[ \dfrac{(\theta-\beta)^2}{2}- \psi \right],
\end{equation}
\begin{equation}
    \delta t=\dfrac{\psi}{c}\,\dfrac{r_sr_d}{r_{ds}}\,\dfrac{n\,\Omega_m^{Real}}{3}\,\dfrac{\mathcal{H}_d}{H_0S_d}\,r_sr_d
\end{equation}
\end{subequations}
We see that $\Delta t_0$ is the usual expression of the time delay, see equation (89) in \cite{bart}, while $\delta t$ is a correction proportional to $\mathcal{H}_d$ and dependent on the distances from the source and the lens. This could be a way to measure this parameter: any difference between the time delay calculated with the usual expression with only baryonic matter included and the measured one could be attributed to this correction, once all systematics have been taken care of. From paper \cite{mio2}, we know that extra dimensions can be the source of what we observe as dark matter, here we find another proof: the term $\mathcal{H}_d/H_d$ affects the apparent mass of the lens changing the time delay, as dark matter would do. We notice that a negative $\mathcal{H}_d$ (contracting dimensions) would increase the time delay, a positive one (expanding dimensions) will decrease it. In \cite{mio}, we have found that $\mathcal{H}<0$ thus the time delay from gravitational lensing is increased, as dark matter would do. So the picture is consistent. As discussed in \cite{tension,td3}, time delay is mostly sensible to the Hubble parameter $H_0$, but it can be used to put constraints on other cosmographic parameters. The recovery of such parameters depends on the reconstruction of the lens potential; here we have found that there is an additional term depending on the expansion rate of the extra dimensions that affects the reconstruction: if one does not take this into consideration, the reconstruction of $\psi$ is biased and the recovery of cosmographic parameters is somewhat invalidated. The likelihood-based reconstruction described in \cite{td3} and references therein should be modified considering the $\psi$ generated by the baryonic mass with the additional and separated contribution due to the expansion rate of the extra dimensions. We shall now show how to modify this procedure in the case of weak lensing.

\section{Mass reconstruction and parameter estimation}
\label{sec:mrec}

It is possible to reconstruct the mass profile of the lens by studying the ellipticity of the lensed galaxies. As discussed in \cite{reco1,reco2} the measured ellipticity $\epsilon^{(m)}$ is linked to the real ellipticity $\epsilon^{(s)}$ by the expression
\begin{equation}
    \epsilon^{m}=\dfrac{\epsilon^{(s)}+g}{1+g^*\epsilon^{(s)}}, \qquad g=\dfrac{\gamma}{1+K}.
\end{equation}
As we have seen, the convergence and the shear have corrections due to the extra dimensions $\gamma=\gamma_0+\delta\gamma$ and $K=K_0+\delta K$, then the corrections to reduced shear are given by
\begin{equation}
    \delta g=\dfrac{\delta\gamma_1\,\gamma_1+\delta_2\gamma_2}{\gamma\,(1+K_0)}-\dfrac{\gamma_0\,\delta K}{(1+K_0)^2}.
\end{equation}
From the definitions of the convergence and of the shear, we have
\begin{align}
    \delta K&=-\dfrac{4\pi G M}{c^2}\,\dfrac{r_{ds}^2r_d^2}{r_s}\,\dfrac{n\,\Omega^{Real}_m}{3}\,\dfrac{\mathcal{H}_d}{H_0S_d}\,\delta(r_d\theta),\\
    \delta\gamma_1&=\dfrac{4\pi G M}{c^2}\,\dfrac{r_{ds}^2r_d^2}{r_s}\,\dfrac{n\,\Omega^{Real}_m}{3}\,\dfrac{\mathcal{H}_d}{H_0S_d}\,\dfrac{\theta_1^2-\theta_2^2}{\theta^4},\\
    \delta\gamma_2&=\dfrac{4\pi G M}{c^2}\,\dfrac{r_{ds}^2r_d^2}{r_s}\,\dfrac{n\,\Omega^{Real}_m}{3}\,\dfrac{\mathcal{H}_d}{H_0S_d}\,\dfrac{\theta_1\theta_2}{\theta^4}
\end{align}
therefore, the correction $\delta g$ is proportional to $\mathcal{H}_d$ and $n$. These corrections are small (in \cite{mio} we estimated that $\mathcal{H}_0/H_0\ll1$) therefore we can expand the above expression for the ellipticity up to the first order in $\mathcal{H}_d/H_0$, obtaining
\begin{equation}
    \epsilon^{(m)}=\epsilon_0^{(m)}+\delta\epsilon^{(m)}=\dfrac{\epsilon^{(s)}+g_0}{1+g_0^*\epsilon^{(s)}}-\left( \dfrac{\delta\gamma_1\,\gamma_1+\delta_2\gamma_2}{\gamma\,(1+K_0)}-\dfrac{\gamma_0\,\delta K}{(1+K_0)^2} \right)\,\dfrac{(\epsilon^{(s)})^2-1}{(1+\epsilon^{(s)}\,g_0)^2}.
\end{equation}
The mean observable ellipticity is an unbiased estimate of the reduced shear $g$ \cite{reco2,reco3,reco4}
\begin{equation}
    \langle \epsilon^{(m)}\rangle=g=g_0+\delta g.
\end{equation}

As discussed in \cite{reco1}, the reconstruction procedure uses the following likelihood (see also \cite{sch}), where we include our corrections:
\begin{equation}\label{eq:reco}
    -\ln\mathcal{L}= \sum_{i=1}^N\left[ \dfrac{\epsilon_i^{(m)}-g_0(\theta_i,\psi_n)-\delta g(\theta_i,\psi_n)}{\sigma_i^2(\theta_i,\psi_n)}+2\ln\sigma_i(\theta_i,\psi_n) + \lambda_eS(\psi_n) \right],
\end{equation}
in the above expression, $N$ is the number of galaxies in the image, $\sigma_i$ are the standard deviations of the ellipticities, $g_0$ is the leading order reduced shear $\delta g$ its correction, $\lambda_e$ is a Lagrange multiplier, $S$ is the effect of the strong lensing and $\psi_n$ is the discretized lens potential. As we have seen, $\delta g$ depends on the Hubble parameter of the extra dimensions $\mathcal{H}_d$ and on $n$ (and also on the in principle unknown $\Omega^{Real}_m$), therefore, keeping it as a parameter, we could find an estimate for it and at the same time reconstruct the mass profile of the lens.

\section{Weak lensing}
\label{sec:weak}

In this Section, we calculate the weak lensing power spectrum and the effects of lensing on CMB following \cite{bart}.

We first Fourier expand the potential $\phi$
\begin{equation}
    \phi(\vec{x})=\int\dfrac{d^3q}{(2\pi)^3}\,\hat{\phi}(\vec{q})\exp(i\,\vec{q}\cdot\vec{x})
\end{equation}
and introduce the power spectrum 
\begin{equation}
    \langle\hat{\phi}(\vec{q}),\hat{\phi}(\vec{q}^\prime)\rangle=(2\pi)^2\delta(\vec{q}-\vec{q}^\prime)\,P_{\phi}(q).
\end{equation}
Then, the angular correlation function of the lensing potential is
\begin{equation}
    \langle\psi(\theta),\psi(\theta^\prime)\rangle=\int_0^{r_s}dr\int_0^{r_s} dr^\prime \dfrac{f(r_s-r)}{f(r_s)f(r)}\,\dfrac{f(r_s-r^\prime)}{f(r_s)f(r^\prime)}\,\int\dfrac{d^3q}{(2\pi)^3}\,P_\phi(q)\,\exp\big( i\, \vec{q}\cdot(\vec{r}-\vec{r}^\prime) \big).
\end{equation}
As above, we expand the terms containing the $f(x)$ kernel up to the first order in $\mathcal{H}_0$, thus obtaining
\begin{equation}
    \langle\psi(\theta),\psi(\theta^\prime)\rangle=\int_0^{r_s}dr\int_0^{r_s} dr^\prime \dfrac{(r_s-r)(r_s-r^\prime)}{r\,r^\prime\,r_s}\,\left( 1-\dfrac{n\,Omega^{Real}_m}{2}\,\dfrac{\mathcal{H}_d}{H_d}\,(r+r^\prime)\,r_s \right)\,\int\dfrac{d^3q}{(2\pi)^3}\,P_\phi(q)\,\exp\big( i\, \vec{q}\cdot(\vec{r}-\vec{r}^\prime) \big)
\end{equation}
moreover we expand the exponential using spherical Bessel functions and spherical harmonics
\begin{equation}
    \exp(i\vec{q}\cdot\vec{x})=4\pi\,\sum_{lm}\,i^l\,j_l(qx)\,Y^*_{lm}(\theta_x)Y_{lm}(\theta_q)
\end{equation}
where $\theta_x$ and $\theta_q$ are the direction angles of $\vec{x}$ and $\vec{q}$. Expanding the lens potential in spherical harmonics
\begin{equation}
    \psi(\theta)=\sum_{lm}\psi_{lm}Y_{lm}
\end{equation}
and defining the angular spectrum
\begin{equation}
    \langle\psi_{lm},\psi_{l^\prime m^\prime}\rangle=\delta_{ll^\prime}\delta_{mm^\prime}\,C^\psi_l,
\end{equation}
we find
\begin{equation}
    C^\psi_l=\dfrac{2}{\pi}\,\int_0^{r_s}dr\int_0^{r_s} dr^\prime \dfrac{(r_s-r)(r_s-r^\prime)}{r\,r^\prime\,r_s}\,\left( 1-\dfrac{n\,Omega^{Real}_m}{3}\,\dfrac{\mathcal{H}_d}{H_0S_d}\,(r+r^\prime)\,r_s \right)\int_0^\infty dq\,q^2\,P_\phi(q)j_l(qr)j_l(qr^\prime).
\end{equation}
following again \cite{bart}, we assume that the dimension of the large scale structure is smaller than the distances involved, so that
\begin{equation}
    \int dq\, q^2 j_l(qr)j_l(qr^\prime)=\dfrac{\pi}{2r^2}\,\delta(r-r^\prime)
\end{equation}
and $P_\phi(q)\approx P_\phi(l/r)$, arriving to
\begin{subequations}
\begin{equation}
    C_l^\psi=\int_0^{r_s}\,dr\,\left(\dfrac{(r_s-r)}{r\,r_s}\right)^2\,\left( 1-\dfrac{2n}{3}\,\dfrac{\mathcal{H}_d}{H_0S_d}\,{\Omega^{Real}_m}\,r\,r_s \right)\,P_\phi\left(\dfrac{l}{r}\right)={}^0C^\psi_l+\delta C^\psi_l,
\end{equation}
where we have imposed
\begin{equation}
    {}^0C^\psi_l=\int_0^{r_s}\,dr\,\left(\dfrac{(r_s-r)}{r\,r_s}\right)^2\,P_\phi\left(\dfrac{l}{r}\right)
\end{equation}
\begin{equation}
    \delta C^\psi_l=-\int_0^{r_s}\,dr\,\left(\dfrac{(r_s-r)}{r\,r_s}\right)^2\,\dfrac{2n}{3}\,\dfrac{\mathcal{H}_d}{H_0S_d}\,{\Omega^{Real}_m}\,r\,r_s \,P_\phi\left(\dfrac{l}{r}\right).
\end{equation}
\end{subequations}
So we find that there is a correction in the angular spectrum due to the extra dimensions. This correction is negative if $\mathcal{H}_d>0$ (expanding extra dimensions) and viceversa, it is positive for $\mathcal{H}_d<0$ (contracting extra dimensions). We notice that the correction $\delta C^\psi_l$ depends on the distance from the source $r_s$.

If we repeat the treatment given in Sections 3.2--3.6 of \cite{bart}, we find the effect of the gravitational lensing on the CMB temperature power spectrum $C^\tau_l$; it is given by
\begin{equation}
    C^\tau_{l,obs}=C^\tau_l\,(1-l^2R_\psi-l^2\,\delta R_\psi)+\int\dfrac{d^2l_1}{(2\pi)^2}\,\left[ \vec{l}_1\cdot(\vec{l}-\vec{l}_1) \right]^2\,C^\psi_{l_1}\,{}^0C^\tau_{|\vec{l}-\vec{l}_1|}+\int\dfrac{d^2l_1}{(2\pi)^2}\,\left[ \vec{l}_1\cdot(\vec{l}-\vec{l}_1) \right]^2\,C^\psi_{l_1}\,\delta C^\tau_{|\vec{l}-\vec{l}_1|},
\end{equation}
where $C^\tau_{l,obs}$ is the observed spectrum, $C^\tau_l$ is the emitted spectrum, and
\begin{equation}
    R_\psi=\dfrac{1}{3\pi}\, \int^\infty_0 dl\,l^3\,{}^0C^\psi_l, \qquad \delta R_\psi=\dfrac{1}{3\pi}\, \int^\infty_0 dl\,l^3\,\delta C^\psi_l.
\end{equation}
Therefore, there are two additional corrections on the temperature power spectrum given by the correction $\delta C^\psi_l$ to the angular spectrum of the lens potential. For the polarization spectrum of the CMB, we find
\begin{align}\nonumber
    C^{\mathcal{E}}_{l,obs}&=C^{\mathcal{E}}_l\,(1-l^2R_\psi-l^2\,\delta R_\psi)+\\
    &+\int \dfrac{d^2l_1}{(2\pi)^2}\,\left[ \vec{l}_1\cdot(\vec{l}-\vec{l}_1) \right]^2\,\cos^2(\phi_1-\phi)\,\left( {}^0C^\psi_{|\vec{l}-\vec{l}_1|}+\delta C^\psi_{|\vec{l}-\vec{l}_1|} \right)\,C^{\mathcal{E}}_{l_1},\\
    C^{\mathcal{B}}_{l,obs}&=\int \dfrac{d^2l_1}{(2\pi)^2}\,\left[ \vec{l}_1\cdot(\vec{l}-\vec{l}_1) \right]^2\,\sin^2(\phi_1-\phi)\,\left( {}^0C^\psi_{|\vec{l}-\vec{l}_1|}+\delta C^\psi_{|\vec{l}-\vec{l}_1|} \right)\,C^{\mathcal{E}}_{l_1}
\end{align}
from which we find, as already known \cite{bmode}, that gravitational lensing creates the $\mathcal{B}$ modes, which should be absent in the CMB spectrum, but they are also induced by the correction to the angular spectrum of the lens potential.

\section{Conclusion}
In this paper, we have analyzed gravitational lensing in light of the model we have developed, in which dark energy and dark matter can be an effect of extra dimensions. We have found that the extra dimensions affect the time delay and the mass reconstruction procedure as dark matter would do; in particular, we have found that expanding extra dimensions would reduce the time delay, while contracting extra dimensions would increase it; at the same time, the convergence and the shear would increase for contracting extra dimensions, as dark matter would do. This picture is consistent with what we found in \cite{mio} in which contracting extra dimensions are necessary to explain the accelerated expansion of the Universe and in \cite{mio2} in which contracting extra dimensions are necessary to explain the growth of cosmological overdensities without dark matter. 

Using the formulae we have derived here, in particular equation \eqref{eq:reco}, one could study the Bullet Cluster and complete the analysis started in \cite{mio2}, in order to verify that we can explain its configuration even without dark matter. Moreover, we notice that with a tomographic or 3D analysis, one can study the evolution of the parameter $\mathcal{H}_d$ throughout the history of the Universe and map the change in the radius of the extra dimensions. However, this needs a rigorous treatment of the redshift-depending corrections in the Green's function in order to be able to treat with our formalism extended lenses, such as clusters.

\bibliography{biblio}

\end{document}